\documentclass[onecolumn,showpacs]{revtex4}

\usepackage{amsmath}
\usepackage{graphicx}
\usepackage{dcolumn}
\usepackage{bm}

\input epsf

\begin{document}

\title{\Large Parameterizing Dark Energy Models and Study of Finite Time Future Singularities}

\author{\bf~Tanwi~Bandyopadhyay$^1$\footnote{tanwi.bandyopadhyay@aiim.ac.in}
and~Ujjal~Debnath$^2$\footnote{ujjaldebnath@gmail.com,
ujjal@associates.iucaa.in}}

\affiliation{$^1$Department of Mathematics, Adani Institute of
Infrastructure Engineering, Ahmedabad-382421, Gujarat, India.\\
$^2$Department of Mathematics, Indian Institute of Engineering
Science and Technology, Shibpur, Howrah-711103, India.}

\begin{abstract}
A review on spatially flat D-dimensional
Friedmann-Robertson-Walker (FRW) model of the universe has been
performed. Some standard parameterizations of the equation of
state parameter of the Dark Energy models are proposed and the
possibilities of finite time future singularities are
investigated. It is found that certain types of these
singularities may appear by tuning some parameters appropriately.
Moreover, for a scalar field theoretic description of the model,
it was found that the model undergoes bouncing solutions in some
favorable cases.\\


\end{abstract}


\maketitle

\section{\normalsize\bf{Introduction}}

In order to avoid the initial singularity problem \cite{Borde}, a
competitive alternative structure is proposed to the standard
inflationary description. This is called the big bounce scenario
(\cite{Novello}-\cite{Odintsov2}). In this framework, the universe
initially contracts continuously up to an initial narrow state
with a non-zero minimal radius and then transforms to an expanding
phase \cite{Cai4}. This means that the initial singularity is
replaced by a bounce and consequently the cosmic equation of state
(EoS) parameter crosses the phantom divide line (from $\omega<-1$
to $\omega>-1$).\\

On the other hand, various cosmological observations indicate the
current cosmic acceleration (\cite{Riess1}-\cite{Abazajian}). To
explain this phenomenon in the homogeneous and isotropic universe,
it is necessary to assume a component of matter with large
negative pressure, called dark energy. Since the inception of this
concept, a wide variety of phenomenological models have been
proposed as the most suitable candidate of dark energy. Among
them, the cosmological constant provides plausible answers but
suffers from the fine-tuning problem. This issue has forced
researchers to probe models having time-dependent equation of
state parameter for the dark energy component. A simple
classification of these models could be as
follows:\\

1. Cosmological constant ($\omega\equiv \frac{p}{\rho}=-1$)\\

2. Dark energy with $\omega=$ constant $\neq -1$ [cosmic strings
($\omega=-\frac{1}{3}$), domain walls ($\omega=-\frac{2}{3}$)].\\

3. Dynamical dark energy ($\omega=\omega(z)\neq$
constant) (i.e., quintessence, Chaplygin gas, k-essence, braneworld) (\cite{Ratra}-\cite{Sahni})\\

4. Dark energy with $\omega(z)<-1$ [scalar-tensor theory, phantom
models] (\cite{Starobinsky},\cite{Cline})\\

The reconstruction problem for dark energy is reviewed in
\cite{Shah0} in much detail (see also the references therein).
Recently, in the literature \cite{Sadatian}, the possibilities of
future singularities have been studied in the framework of
Modified Chaplygin Gas filled universe and conditions of bounce
were investigated. In this work, we provide more general results
in terms of the possibilities of finite time future singularities
in the context of a D-dimensional Friedmann-Robertson-Walker
universe by imposing five standard parameterizations of the dark
energy equation of state parameter. Different aspects of dark
energy models have been studied so far to reconcile standard
cosmological scenario with observations but no work has been done
to the parameterization of these models in view of investigating
the finite time future singularities.\\

The paper is organized as follows. In section II, we state the
governing equations of the metric in the D-dimensional Universe.
In section III, the description for the major physical quantities
are provided in the backdrop of five different parameterization
models both analytically and graphically. In section IV, the
possibilities of future singularities are examined for these five
models and the possible restrictions for certain type of
singularity are shown in tabular form. In Section V, the bouncing
universe is described in the five parameterization model and
finally, we end with a short
discussion in section VI.\\

\section{\normalsize\bf{D-dimensional Friedmann-Robertson-Walker Universe}}

We consider the D-dimensional spatially flat
Friedmann-Robertson-Walker universe, given by \cite{Sadatian}

\begin {equation}
ds^2=-dt^2+{a^2}(t)d{\Omega}^2
\end{equation}

where $a(t)$ is the scale factor and $d{\Omega}^2$ is the metric
for the $(D-1)$ dimensional space. The field equations together
with the energy conservation equation can be obtained as (assuming
$8\pi G=c=1$)

\begin{equation}\label{fld eqn 1}
H^2=\frac{2\rho}{(D-1)(D-2)}
\end{equation}

\begin{equation}
\dot{H}+H^2=-\frac{(D-1)p+(D-3)\rho}{(D-1)(D-2)}
\end{equation}

and

\begin{equation}
\dot{\rho}+(D-1)H(\rho+p)=0
\end{equation}

Here $H=\frac{\dot{a}}{a}$ denotes the Hubble parameter. In the
following section and subsequent sub-sections, we will examine
different parameterizations of dark energy models and attain
analytical as well as graphical expressions of some
physical entities involved in the process.\\

\section{\normalsize\bf{Dark Energy as Scalar Field for Various Parameterization Models}}

By considering scalar field $\phi$ with the potential $U(\phi)$,
the effective Lagrangian \cite{Sadatian} is given by

\begin{equation}
L_\phi=\frac{{\dot{\phi}}^2}{2}-U
\end{equation}

So the energy density and pressure take the form

\begin{equation}
\rho=\frac{{\dot{\phi}}^2}{2}+U
\end{equation}

\begin{equation}
p=\frac{{\dot{\phi}}^2}{2}-U
\end{equation}

The kinetic energy for the field is given by

\begin{equation}
{\dot{\phi}}^2=(1+\omega)\rho
\end{equation}

where $\omega=\frac{p}{\rho}$. One can always write
$\dot{\phi}={\phi}'\dot{z}$, where the prime denotes
differentiation with respect to redshift $z=\frac{1}{1+a}$. With
the help of \eqref{fld eqn 1}, we have

\begin{equation}\label{phi derivative}
{\phi}'=\sqrt{\frac{(D-1)(D-2)(1+\omega)}{2}}\frac{1}{1+z}
\end{equation}

The scalar potential associated with the field is given by

\begin{equation}\label{potential}
U=\frac{1}{2}(1-\omega)\rho
\end{equation}

In the next subsections, we will investigate the scalar field and
its potential in different well known parameterization models. A
detailed review of different dark energy models can be found in
\cite{Bamba}.

\subsection{\normalsize\bf{Model I : Linear Parameterization}}

Here we assume the linear equation of state \cite{Coor}
$\omega=\omega(z)=\omega_0+{\omega_1}z$. Here $\omega_0$ and
$\omega_1$ are constants. Using equation (4), the energy density
of the model then gives rise to

\begin{equation}
\rho=\rho_0(1+z)^{(D-1)(1+\omega_0-{\omega_1})}e^{(D-1){\omega_1}z}
\end{equation}

Here $\rho_0$ is the present value of the energy density. The
pressure of the fluid then becomes

\begin{equation}
p=(\omega_0+{\omega_1}z)\rho_0(1+z)^{(D-1)(1+\omega_0-{\omega_1})}e^{(D-1){\omega_1}z}
\end{equation}

Then \eqref{phi derivative} reduce to

\begin{equation}
{\phi}'=\sqrt{\frac{(D-1)(D-2)}{2}}\frac{\sqrt{1+\omega_0+{\omega_1}z}}{1+z}
\end{equation}

which on further integration gives rise to the scalar field as

\begin{equation}
\phi=\phi_0+\sqrt{2(D-1)(D-2)}\left[\sqrt{1+\omega_0+{\omega_1}z}-M
tan^{-1}\left(\frac{\sqrt{1+\omega_0+{\omega_1}z}}{M}\right)\right]
\end{equation}

where $M=\sqrt{\omega_1-\omega_0-1}$. Also \eqref{potential}
changes to

\begin{equation}
U=\frac{1}{2}(1-\omega_0-{\omega_1}z)\rho_0(1+z)^{(D-1)(1+\omega_0-{\omega_1})}e^{(D-1){\omega_1}z}
\end{equation}

We draw the variations of energy density $\rho$, pressure $p$, EoS
parameter $\omega(z)$, Hubble parameter $H$, potential $U(\phi)$
and the scalar field $\phi$ with the variation in $z$ in Figs. 1 -
6 respectively. Fig. 7 shows the variation of $U(\phi)$ with
$\phi$. We observe that $\rho$, $H$ and $\phi$ decrease as $z$
decreases. $p$ and $\omega$ decrease from positive level to
negative level as $z$ decreases. $U(\phi)$ first increases up to
certain value of $z$ and then decreases as $z$ decreases. Also
$U(\phi)$ decreases as $\phi$ increases.\\

\begin{figure}
\includegraphics[height=2.0in]{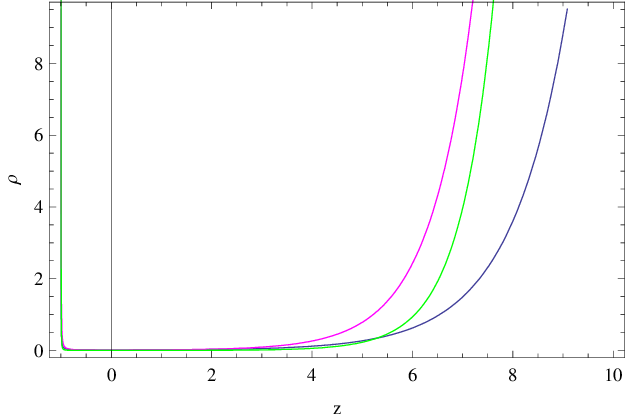}~~~~
\includegraphics[height=2.0in]{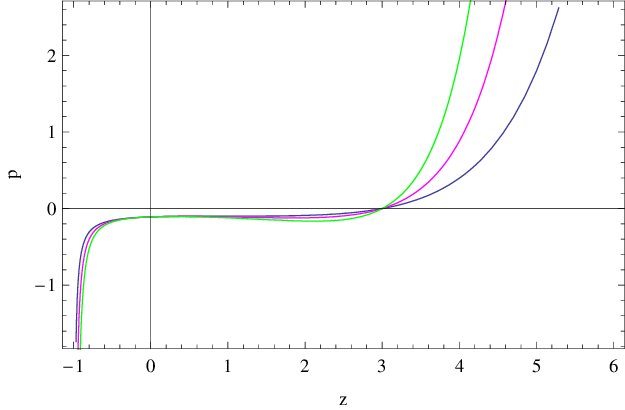}\\

\vspace{2mm} ~~~~~~~~~~~~Fig.1~~~~~~~~~~~~~~~~~~~~~~~~~~~~~~~~~~~~~~~~~~~Fig.2~~~~\\
\vspace{4mm}

\includegraphics[height=2.0in]{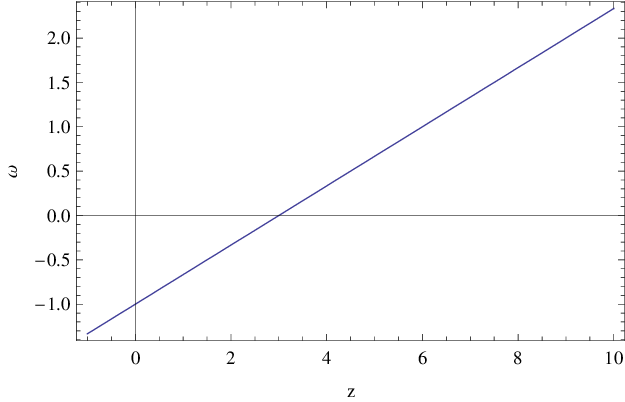}~~~~
\includegraphics[height=2.0in]{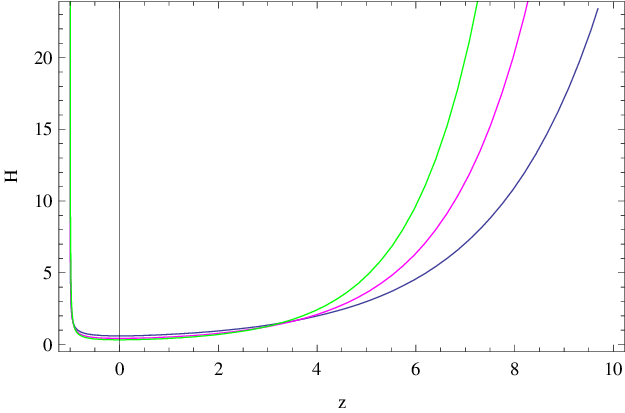}\\

\vspace{2mm} ~~~~~~~~~~~~Fig.3~~~~~~~~~~~~~~~~~~~~~~~~~~~~~~~~~~~~~~~~~~~Fig.4~~~~\\
\vspace{4mm}

\includegraphics[height=2.0in]{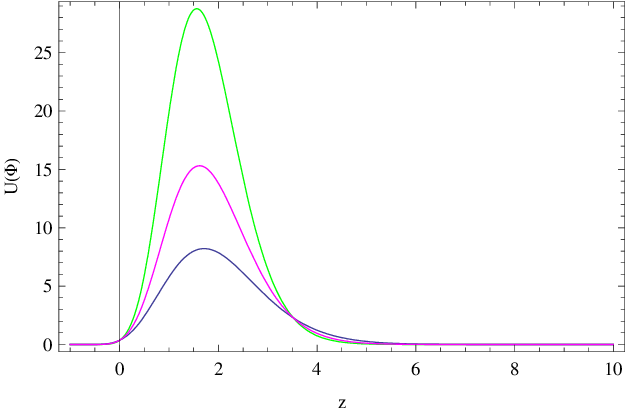}~~~~
\includegraphics[height=2.0in]{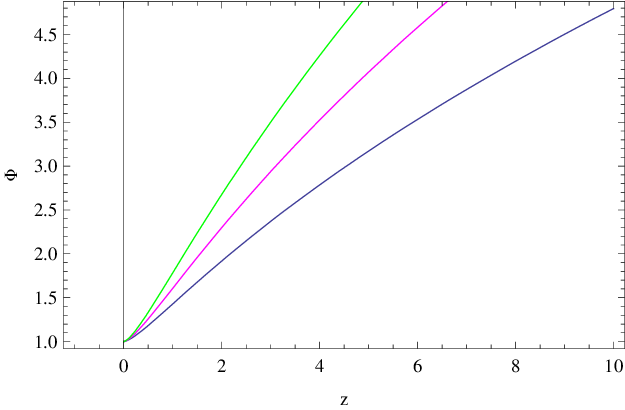}\\

\vspace{2mm} ~~~~~~~~~~~~Fig.5~~~~~~~~~~~~~~~~~~~~~~~~~~~~~~~~~~~~~~~~~~~Fig.6~~~~\\
\vspace{4mm}

\includegraphics[height=2.0in]{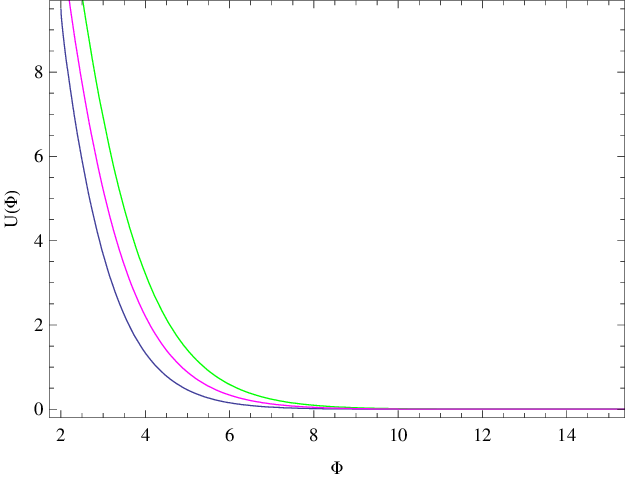}\\

\vspace{2mm} ~~~~~~~~~~~~Fig.7\\

Figs. 1, 2, 3, 4, 5 and 6 show the variation of energy density
$\rho$, pressure $p$, EoS parameter $\omega(z)$, Hubble parameter
$H$, potential $U(\phi)$ and the scalar field $\phi$ with the
variation in $z$ for the {\bf Linear} parameterizations. Fig. 7
shows the variation of $U(\phi)$ with $\phi$. Different colored
paths are obtained for different values of D.\\

\vspace{6mm}

\end{figure}

\subsection{\normalsize\bf{Model II : Chevallier-Polarski-Linder (CPL) Parameterization}}

In the Chevallier-Polarski-Linder (CPL) Parameterization model,
the equation of state is given by \cite{Chev,Linder}
$\omega=\omega(z)=\omega_0+{\omega_1}\frac{z}{1+z}$. Here again
$\omega_0$ and $\omega_1$ are constants. With these, the
expressions of energy density, pressure, scalar field and the
potential become

\begin{equation}
\rho=\rho_0(1+z)^{(D-1)(1+\omega_0+{\omega_1})}e^{-\frac{(D-1){\omega_1}z}{1+z}}
\end{equation}

\begin{equation}
p=\left(\omega_0+{\omega_1}\frac{z}{1+z}\right)\rho_0(1+z)^{(D-1)(1+\omega_0+{\omega_1})}e^{-\frac{(D-1){\omega_1}z}{1+z}}
\end{equation}

\begin{eqnarray*}
\phi=\phi_0+\sqrt{2(D-1)(D-2)}\left[\sqrt{1+\omega_0+\omega_1}~log\left\{(1+\omega_0+\omega_1)\sqrt{1+z}\right.\right.
\end{eqnarray*}
\begin{equation}
\left.\left.+\sqrt{1+\omega_0+\omega_1}\sqrt{(1+\omega_0)(1+z)+{\omega_1}z}\right\}-\sqrt{(1+\omega_0)+\frac{{\omega_1}z}{1+z}}~\right]
\end{equation}
and
\begin{equation}
U=\frac{1}{2}\left(1-\omega_0-{\omega_1}\frac{z}{1+z}\right)\rho_0(1+z)^{(D-1)(1+\omega_0+{\omega_1})}e^{-\frac{(D-1){\omega_1}z}{1+z}}
\end{equation}

We draw the variations of energy density $\rho$, pressure $p$, EoS
parameter $\omega(z)$, Hubble parameter $H$, potential $U(\phi)$
and the scalar field $\phi$ with the variation in $z$ in Figs. 8 -
13 respectively. Fig. 14 shows the variation of $U(\phi)$ with
$\phi$. We observe that $\rho$, $H$, and $U(\phi)$ first decrease
up to $z=0$ as $z$ decreases and then sharply increase as $z$
decreases. But $p$ first increases up to $z=0$ as $z$ decreases
and then sharply decreases as $z$ decreases but keeps the negative
level. $\omega$ decrease as $z$ decreases and always keeps
negative sign. $\phi$ increases as $z$ decreases.
Also $U(\phi)$ decreases as $\phi$ increases.\\

\begin{figure}
\includegraphics[height=2.0in]{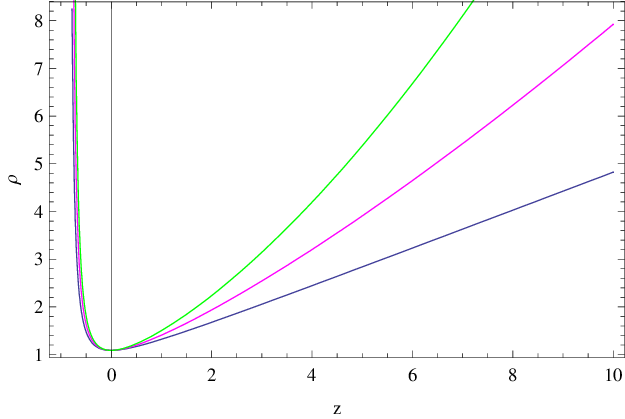}~~~~
\includegraphics[height=2.0in]{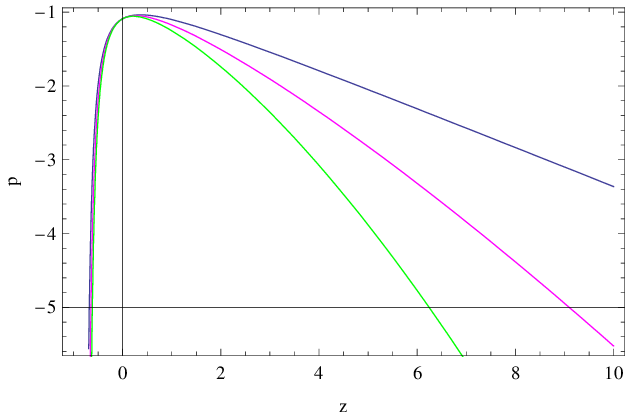}\\

\vspace{2mm} ~~~~~~~~~~~~Fig.8~~~~~~~~~~~~~~~~~~~~~~~~~~~~~~~~~~~~~~~~~~~Fig.9~~~~\\
\vspace{4mm}

\includegraphics[height=2.0in]{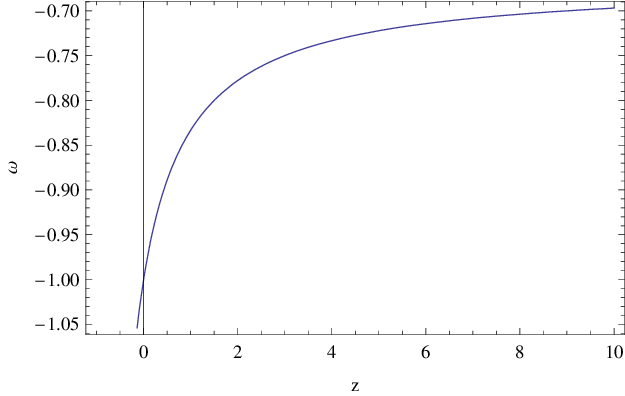}~~~~
\includegraphics[height=2.0in]{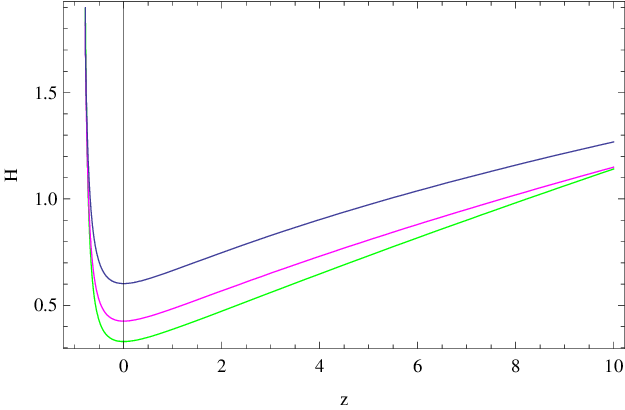}\\

\vspace{2mm} ~~~~~~~~~~~~Fig.10~~~~~~~~~~~~~~~~~~~~~~~~~~~~~~~~~~~~~~~~~~~Fig.11~~~~\\
\vspace{4mm}

\includegraphics[height=2.0in]{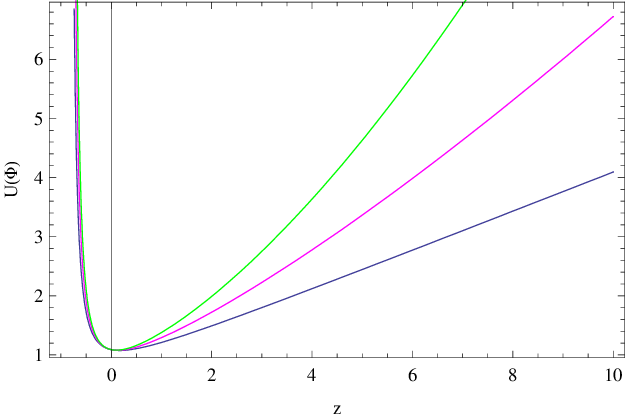}~~~~
\includegraphics[height=2.0in]{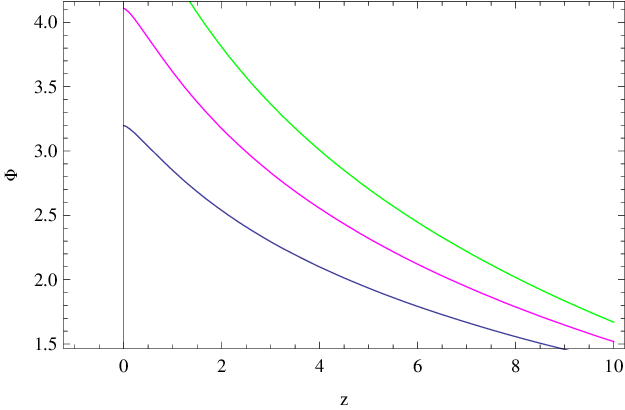}\\

\vspace{2mm} ~~~~~~~~~~~~Fig.12~~~~~~~~~~~~~~~~~~~~~~~~~~~~~~~~~~~~~~~~~~~Fig.13~~~~\\
\vspace{4mm}

\includegraphics[height=2.0in]{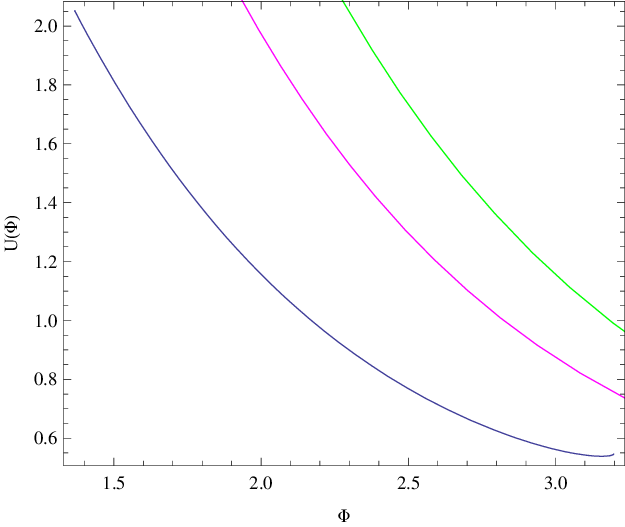}\\

\vspace{2mm} ~~~~~~~~~~~~Fig.14\\

Figs. 8, 9, 10, 11, 12 and 13 show the variation of energy density
$\rho$, pressure $p$, EoS parameter $\omega(z)$, Hubble parameter
$H$, potential $U(\phi)$ and the scalar field $\phi$ with the
variation in $z$ for the {\bf CPL} parameterizations. Fig. 14
shows the variation of $U(\phi)$ with $\phi$. Different colored
paths are obtained for different values of D.\\

\vspace{6mm}

\end{figure}

\subsection{\normalsize\bf{Model III : Jassal-Bagla-Padmanabhan (JBP) Parameterization}}

For the Jassal-Bagla-Padmanabhan (JBP) Parameterization model, the
equation of state changes to \cite{Jassal}
$\omega=\omega(z)=\omega_0+{\omega_1}\frac{z}{(1+z)^2}$, where
$\omega_0$ and $\omega_1$ are constants. The following expressions
are subsequently obtained as

\begin{equation}
\rho=\rho_0(1+z)^{(D-1)(1+\omega_0)}e^{\frac{(D-1){\omega_1}z^2}{2(1+z)^2}}
\end{equation}

\begin{equation}
p=\left[\omega_0+{\omega_1}\frac{z}{(1+z)^2}\right]\rho_0(1+z)^{(D-1)(1+\omega_0)}e^{\frac{(D-1){\omega_1}z^2}{2(1+z)^2}}
\end{equation}

\begin{eqnarray*}
\phi=\phi_0+\sqrt{\frac{(D-1)(D-2)}{2}}\left[-\sqrt{1+\omega_0+\frac{{\omega_1}z}{(1+z)^2}}+\frac{\sqrt{\omega_1}}{2}
~tan^{-1}\left\{\frac{\sqrt{\omega_1}(z-1)}{\sqrt{{\omega_1}z+(1+\omega_0)(1+z)^2}}\right\}\right.
\end{eqnarray*}
\begin{equation}
\left.+\sqrt{1+\omega_0}~Log\left\{\omega_1+2(1+\omega_0)(1+z)+2\sqrt{1+\omega_0}\sqrt{{\omega_1}z+(1+\omega_0)(1+z)^2}\right\}\right]
\end{equation}

\begin{equation}
U=\frac{1}{2}\left[1-\omega_0-{\omega_1}\frac{z}{(1+z)^2}\right]\rho_0(1+z)^{(D-1)(1+\omega_0)}e^{\frac{(D-1){\omega_1}z^2}{2(1+z)^2}}
\end{equation}

We draw the variations of energy density $\rho$, pressure $p$, EoS
parameter $\omega(z)$, Hubble parameter $H$, potential $U(\phi)$
and the scalar field $\phi$ with the variation in $z$ in Figs. 15
- 20 respectively. Fig. 21 shows the variation of $U(\phi)$ with
$\phi$. We observe that $\rho$, $H$, and $U(\phi)$ first decrease
up to $z=0$ as $z$ decreases and then sharply increase as $z$
decreases. But $p$ first increases up to $z=0$ as $z$ decreases
and then sharply decreases as $z$ decreases but keeps the negative
level. $\omega$ first increases up to certain value of $z$ and
then sharply decreases but keeps negative sign. $\phi$ decreases
as $z$ decreases. Also $U(\phi)$ increases as $\phi$ increases.\\

\begin{figure}
\includegraphics[height=2.0in]{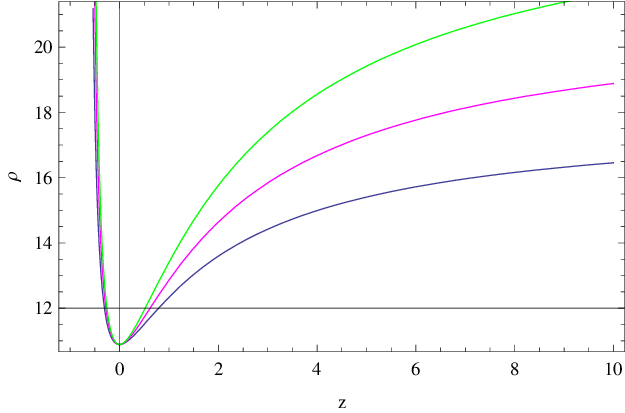}~~~~
\includegraphics[height=2.0in]{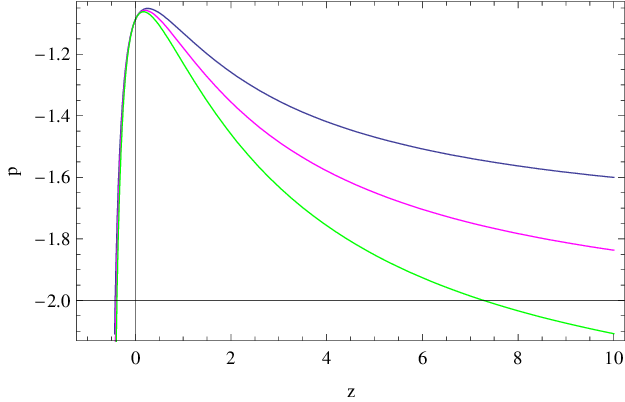}\\

\vspace{2mm} ~~~~~~~~~~~~Fig.15~~~~~~~~~~~~~~~~~~~~~~~~~~~~~~~~~~~~~~~~~~~Fig.16~~~~\\
\vspace{4mm}

\includegraphics[height=2.0in]{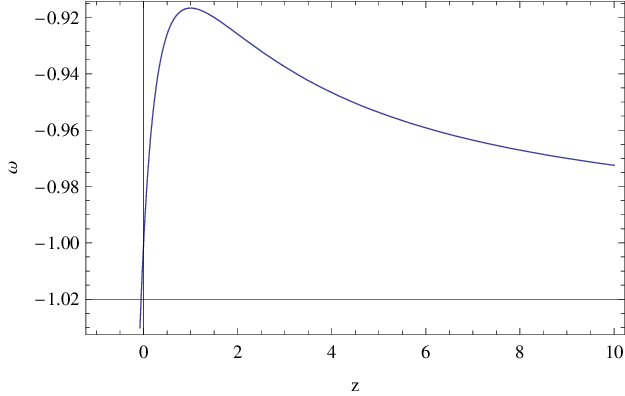}~~~~
\includegraphics[height=2.0in]{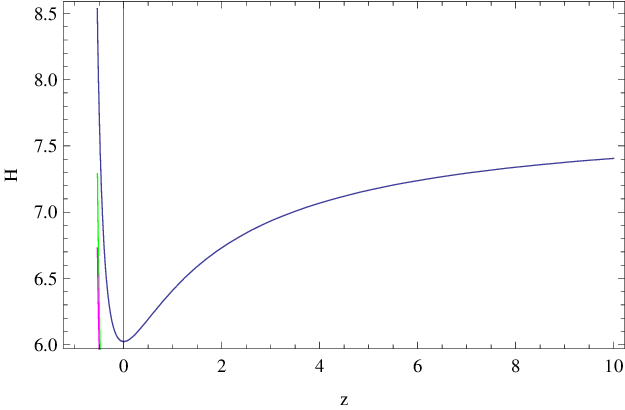}\\

\vspace{2mm} ~~~~~~~~~~~~Fig.17~~~~~~~~~~~~~~~~~~~~~~~~~~~~~~~~~~~~~~~~~~~Fig.18~~~~\\
\vspace{4mm}

\includegraphics[height=2.0in]{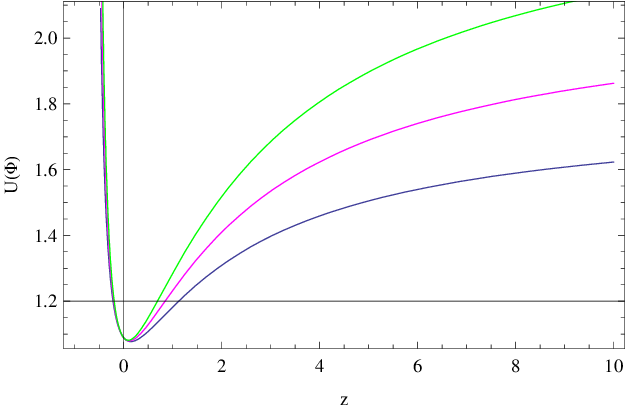}~~~~
\includegraphics[height=2.0in]{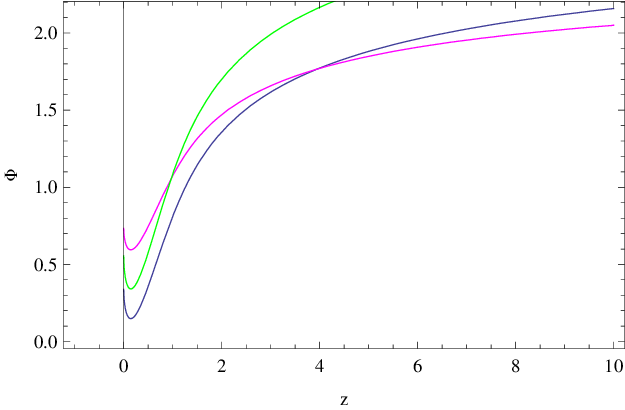}\\

\vspace{2mm} ~~~~~~~~~~~~Fig.19~~~~~~~~~~~~~~~~~~~~~~~~~~~~~~~~~~~~~~~~~~~Fig.20~~~~\\
\vspace{4mm}

\includegraphics[height=2.0in]{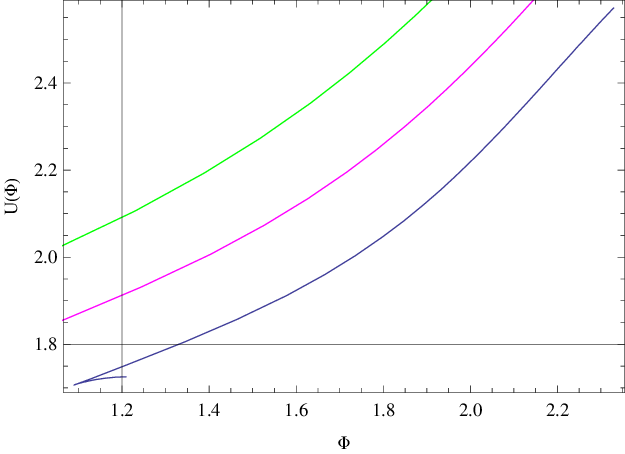}\\

\vspace{2mm} ~~~~~~~~~~~~Fig.21\\

Figs. 15, 16, 17, 18, 19 and 20 show the variation of energy
density $\rho$, pressure $p$, EoS parameter $\omega(z)$, Hubble
parameter $H$, potential $U(\phi)$ and the scalar field $\phi$
with the variation in $z$ for the {\bf JBP} parameterizations.
Fig. 21 shows the variation of $U(\phi)$ with $\phi$. Different
colored paths are obtained for different values of D.\\

\vspace{6mm}

\end{figure}

\subsection{\normalsize\bf{Model IV: Alam-Sahni-Saini-Starobinsky (ASSS) Parameterization}}

In the Alam-Sahni-Saini-Starobinsky (ASSS) parameterization model,
the equation of state parameter has the expression
\cite{Alam,Alam1}

\begin{equation}
\omega=\omega(z)=-1+\frac{A_1(1+z)+2A_2(1+z)^2}{3[A_0+A_1(1+z)+A_2(1+z)^2]}
\end{equation}

where, $A_0,~A_1$ and $A_2$ are constants. In this case, the
physical parameters $\rho,~p,~\phi$ and $U$ become

\begin{equation}
\rho=\rho_0\left[\frac{A_0+A_1(1+z)+A_2(1+z)^2}{A_0+A_1+A_2}\right]^{\frac{D-1}{3}}
\end{equation}

\begin{equation}
p=\left[-1+\frac{A_1(1+z)+2A_2(1+z)^2}{3\left\{A_0+A_1(1+z)+2A_2(1+z)^2\right\}}\right]
~\rho_0\left[\frac{A_0+A_1(1+z)+A_2(1+z)^2}{A_0+A_1+A_2}\right]^{\frac{D-1}{3}}
\end{equation}

\begin{equation}
\phi=\phi_0+\sqrt{\frac{(D-1)(D-2)}{6}}\int\sqrt{\frac{2A_2+\frac{A_1}{1+z}}{A_0+A_1(1+z)+{A_2}(1+z)^2}}dz
\end{equation}

\begin{equation}
U=\left[1-\frac{A_1(1+z)+2A_2(1+z)^2}{6\left\{A_0+A_1(1+z)+A_2(1+z)^2\right\}}\right]
~\rho_0\left[\frac{A_0+A_1(1+z)+A_2(1+z)^2}{A_0+A_1+A_2}\right]^{\frac{D-1}{3}}
\end{equation}

We draw the variations of energy density $\rho$, pressure $p$, EoS
parameter $\omega(z)$, Hubble parameter $H$, potential $U(\phi)$
and the scalar field $\phi$ with the variation in $z$ in Figs. 22
- 27 respectively. Fig. 28 shows the variation of $U(\phi)$ with
$\phi$. We observe that $\rho$, $H$, $\phi$ and $U(\phi)$ decrease
as $z$ decreases. $p$ increases as $z$ decreases but keeps
negative sign. $\omega$ first decreases up to certain value of $z$
and then increases as $z$ decreases. Also $U(\phi)$ increases as $\phi$ increases.\\

\begin{figure}
\includegraphics[height=2.0in]{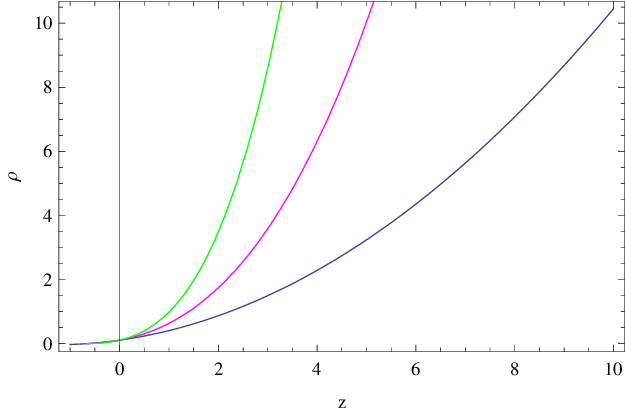}~~~~
\includegraphics[height=2.0in]{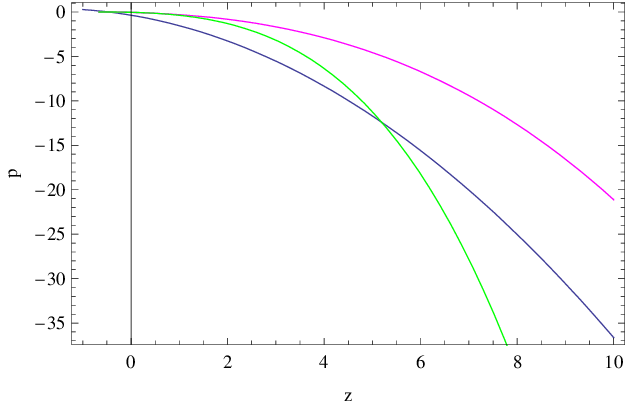}\\

\vspace{2mm} ~~~~~~~~~~~~Fig.22~~~~~~~~~~~~~~~~~~~~~~~~~~~~~~~~~~~~~~~~~~~Fig.23~~~~\\
\vspace{4mm}

\includegraphics[height=2.0in]{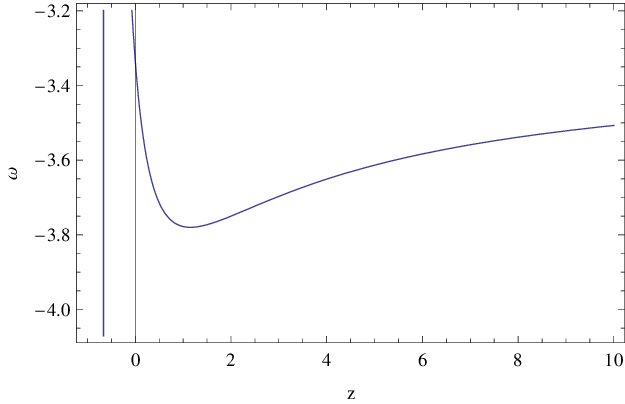}~~~~
\includegraphics[height=2.0in]{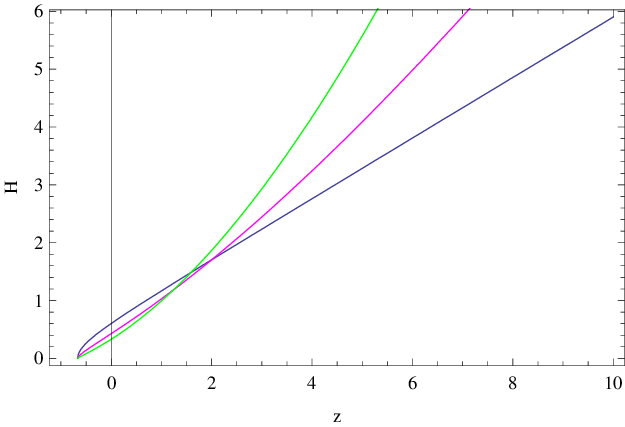}\\

\vspace{2mm} ~~~~~~~~~~~~Fig.24~~~~~~~~~~~~~~~~~~~~~~~~~~~~~~~~~~~~~~~~~~~Fig.25~~~~\\
\vspace{4mm}

\includegraphics[height=2.0in]{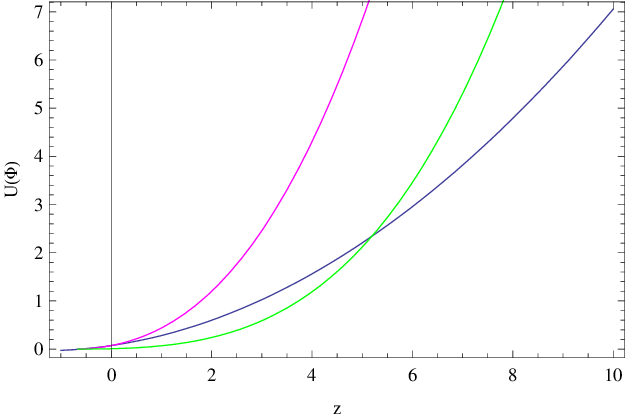}~~~~
\includegraphics[height=2.0in]{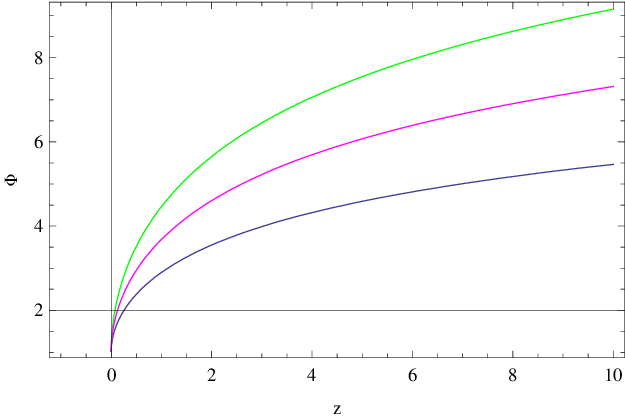}\\

\vspace{2mm} ~~~~~~~~~~~~Fig.26~~~~~~~~~~~~~~~~~~~~~~~~~~~~~~~~~~~~~~~~~~~Fig.27~~~~\\
\vspace{4mm}

\includegraphics[height=2.0in]{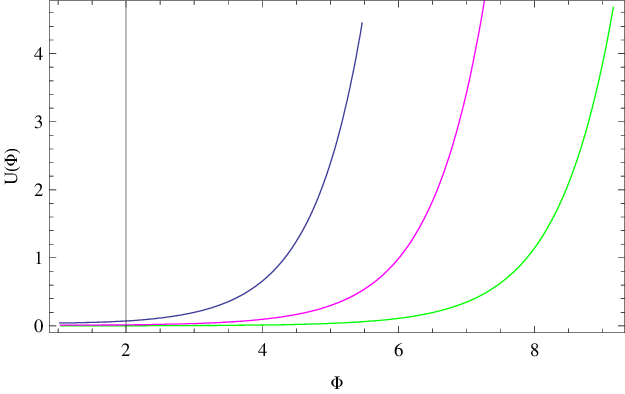}\\

\vspace{2mm} ~~~~~~~~~~~~Fig.28\\

Figs. 22, 23, 24, 25, 26 and 27 show the variation of energy
density $\rho$, pressure $p$, EoS parameter $\omega(z)$, Hubble
parameter $H$, potential $U(\phi)$ and the scalar field $\phi$
with the variation in $z$ for the {\bf ASSS} parameterizations.
Fig. 28 shows the variation of $U(\phi)$ with
$\phi$. Different colored paths are obtained for different values of D.\\

\vspace{6mm}

\end{figure}

\subsection{\normalsize\bf{Model V: Efstathiou
parametrization}}

Here, in Efstathiou parametrization model, the equation of state
takes the form \cite{Ef,Sil}
$\omega=\omega(z)=\omega_0+\omega_1~log(1+z)$, where again
$\omega_0$ and $\omega_1$ are constants. This gives rise to the
following terms

\begin{equation}
\rho=\rho_0(1+z)^{(D-1)(1+\omega_0)}e^{\frac{(D-1)\omega_1}{2}[log(1+z)]^2}
\end{equation}

\begin{equation}
p=[\omega_0+\omega_1~log(1+z)]\rho_0(1+z)^{(D-1)(1+\omega_0)}e^{\frac{(D-1)\omega_1}{2}[log(1+z)]^2}
\end{equation}

\begin{equation}
\phi=\phi_0+\frac{\sqrt{2(D-1)(D-2)}}{3\omega_1}[1+\omega_0+\omega_1~log(1+z)]^{\frac{3}{2}}
\end{equation}

\begin{equation}
U=\frac{1}{2}[1-\omega_0-\omega_1~log(1+z)]\rho_0(1+z)^{(D-1)(1+\omega_0)}e^{\frac{(D-1)\omega_1}{2}[log(1+z)]^2}
\end{equation}

We draw the variations of energy density $\rho$, pressure $p$, EoS
parameter $\omega(z)$, Hubble parameter $H$, potential $U(\phi)$
and the scalar field $\phi$ with the variation in $z$ in Figs. 29
- 34 respectively. Fig. 35 shows the variation of $U(\phi)$ with
$\phi$. We observe that $\rho$, $H$, and $U(\phi)$ first decrease
up to certain value of $z$ as $z$ decreases and then sharply
increase as $z$ decreases. But $p$ first decreases from positive
level to negative level then slightly increases and then sharply
decreases as $z$ decreases but keeps the negative sign. $\omega$
decreases from positive level to negative level as $z$ decreases.
$\phi$ decreases as $z$ decreases. Also $U(\phi)$ increases as $\phi$ increases.\\

\begin{figure}
\includegraphics[height=2.0in]{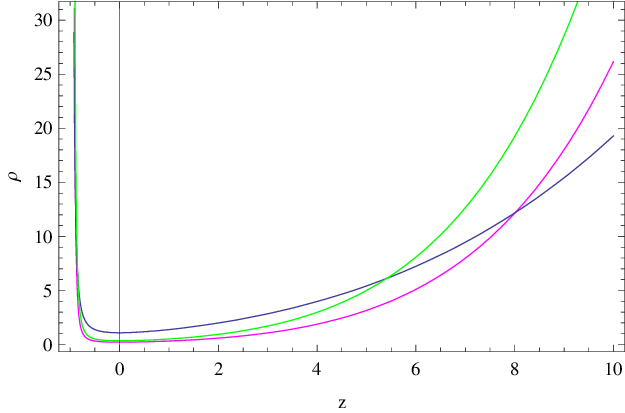}~~~~
\includegraphics[height=2.0in]{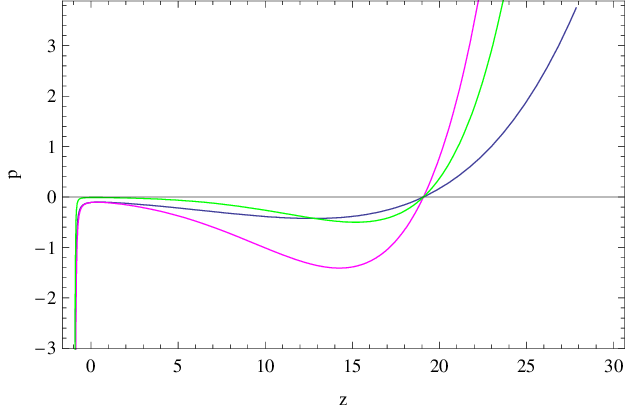}\\

\vspace{2mm} ~~~~~~~~~~~~Fig.29~~~~~~~~~~~~~~~~~~~~~~~~~~~~~~~~~~~~~~~~~~~Fig.30~~~~\\
\vspace{4mm}

\includegraphics[height=2.0in]{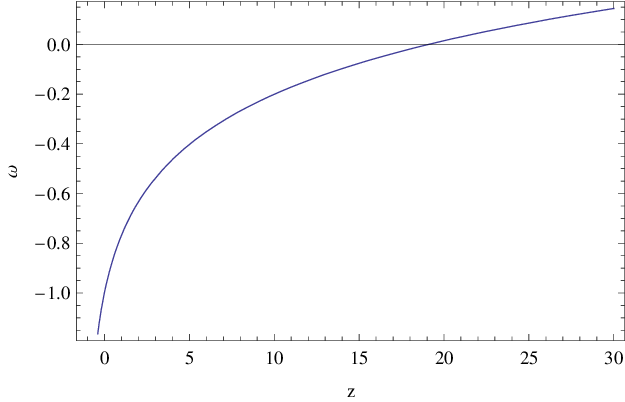}~~~~
\includegraphics[height=2.0in]{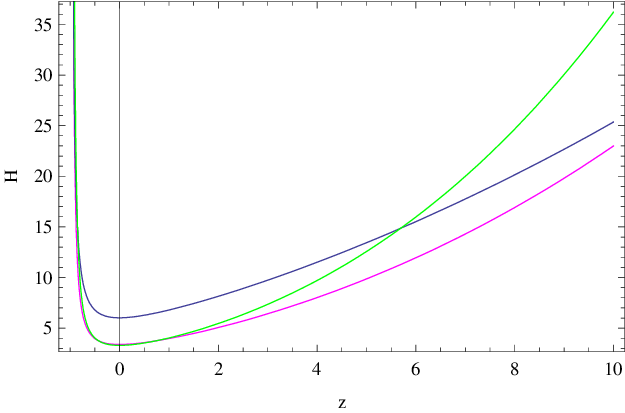}\\

\vspace{2mm} ~~~~~~~~~~~~Fig.31~~~~~~~~~~~~~~~~~~~~~~~~~~~~~~~~~~~~~~~~~~~Fig.32~~~~\\
\vspace{4mm}

\includegraphics[height=2.0in]{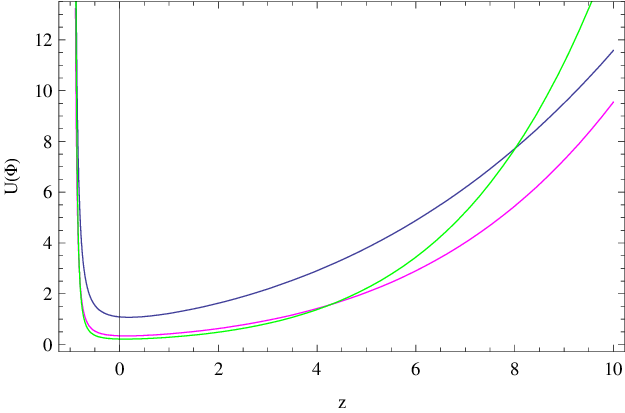}~~~~
\includegraphics[height=2.0in]{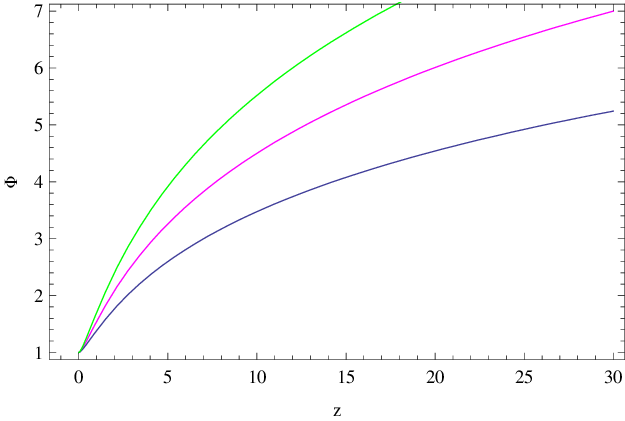}\\

\vspace{2mm} ~~~~~~~~~~~~Fig.33~~~~~~~~~~~~~~~~~~~~~~~~~~~~~~~~~~~~~~~~~~~Fig.34~~~~\\
\vspace{4mm}

\includegraphics[height=2.0in]{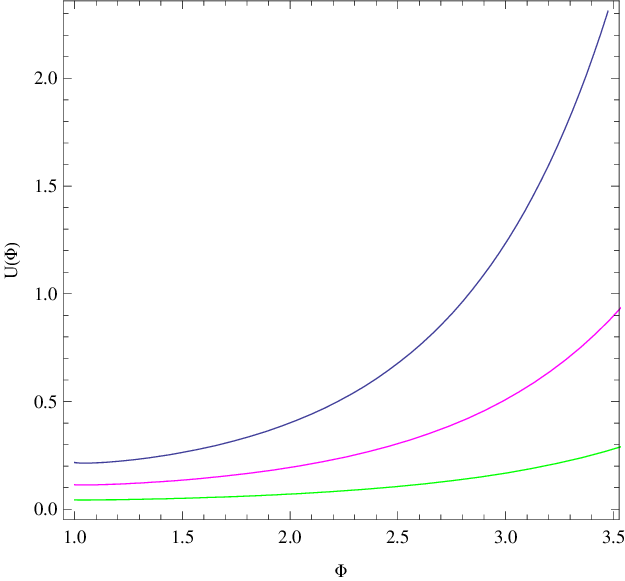}\\

\vspace{2mm} ~~~~~~~~~~~~Fig.35\\

Figs. 29, 30, 31, 32, 33 and 34 show the variation of energy
density $\rho$, pressure $p$, EoS parameter $\omega(z)$, Hubble
parameter $H$, potential $U(\phi)$ and the scalar field $\phi$
with the variation in $z$ for the {\bf Efstathiou}
parameterizations. Fig. 35 shows the variation of $U(\phi)$ with
$\phi$. Different colored paths are obtained for different values of D.\\

\vspace{6mm}

\end{figure}

\section{\normalsize\bf{Analysis of Future Singularities}}

The future singularities can be classified in the following ways
\cite{McInnes,Nojiri}:\\

$\bullet$ Type I (Big Rip) : For $t\rightarrow t_s$, $a\rightarrow
\infty$, $\rho\rightarrow \infty$ and $|p|\rightarrow \infty$\\

$\bullet$ Type II (Sudden) : For $t\rightarrow t_s$, $a\rightarrow
a_s$, $\rho\rightarrow \rho_s$ and $|p|\rightarrow \infty$\\

$\bullet$ Type III : $t\rightarrow t_s$, $a\rightarrow
a_s$, $\rho\rightarrow \infty$ and $|p|\rightarrow \infty$\\

$\bullet$ Type IV : For $t\rightarrow t_s$, $a\rightarrow
a_s$, $\rho\rightarrow 0$ and $|p|\rightarrow 0$\\

where $t_s$, $a_s$ and $\rho_s$ are constants with $a_s\neq 0$.\\

The following table shows the restrictions on the parameters
involved in the five parameterization models for the occurrence of
the future singularities:\\

\begin{center}
{\bf \text TABLE} \\\vspace{.5cm}
\begin{tabular}{|c|c|c|c|c|c|}
  \hline
  ~~~&~~~&~~~&~~~&~~~&\\
  ~~~\textbf{Singularity/Model}~~~ & ~~~\textbf{Linear}~~~ & ~~~\textbf{CPL}~~~ & ~~~\textbf{JBP}~~~ & ~~~\textbf{ASSS}~~~ &
  ~~~\textbf{Efstathiou}~~~\\
  ~~~&~~~&~~~&~~~&~~~&\\
  \hline\hline
  ~~~&~~~&~~~&~~~&~~~&\\
  ~~~\textbf{Type I (Big Rip)}~~~ & ~~~$1+\omega_0-\omega_1<0$~~~ & ~~~$\omega_1>0$~~~ & ~~~$\omega_1>0$~~~ & ~~~No~~~ & ~~~$\omega_1>0$~~~ \\
  ~~~&~~~&~~~&~~~&~~~&\\
  \hline
  ~~~&~~~&~~~&~~~&~~~&\\
  ~~~\textbf{Type II (Sudden)}~~~ & ~~~No~~~ & ~~~No~~~ & ~~~No~~~ & ~~~No~~~ & ~~~No~~~ \\
  ~~~&~~~&~~~&~~~&~~~&\\
  \hline
  ~~~&~~~&~~~&~~~&~~~&\\
  ~~~\textbf{Type III}~~~ & ~~~No~~~ & ~~~No~~~ & ~~~No~~~ & ~~~No~~~ & ~~~No~~~ \\
  ~~~&~~~&~~~&~~~&~~~&\\
  \hline
  ~~~&~~~&~~~&~~~&~~~&\\
  ~~~\textbf{Type IV}~~~ & ~~~No~~~ & ~~~No~~~ & ~~~No~~~ & ~~~${A_1}^2\geq 4A_0A_2$~~~ & ~~~No~~~ \\
  ~~~&~~~&~~~&~~~&~~~&\\
  \hline
\end{tabular}

\end{center}

\section{\normalsize\bf{Bouncing Universe}}

The initial singularity in the cosmological models can be avoided
by introducing the non-singular bouncing models. In these models,
an universe undergoing a 'bounce' stage attains a minimum after a
collapsing phase and then subsequently expands. During the
collapse, the scale factor $a(t)$ decreases [$\dot{a}(t) < 0$] and
during the expanding phase, it increases [$\dot{a}(t) > 0$]. At
the bounce point, $t = t_{b}$, the minimal necessary condition
(may not be sufficient) is $\dot{a}(t)=0$ and $\ddot{a}(t) > 0$
for $t\in (t_{b}-c, t_{b})\cup (t_{b},t_{b}+c)$ for small $c>0$.
For a non-singular bounce $a(t_{b})\ne 0$.\\

In the current study, the bouncing universe can be viewed in
different parameterizing models from the figures 36-45, where the
evolution of the scale factor $a(t)$ and the EoS parameter
$\omega$ are shown with respect to time $t$ for each model. For
the bouncing phase, the scale factor first decreases and then
increases in order that the universe enters into the hot Big Bang
era immediately after the bounce. The nature of the equation of
state parameters have been presented in the figures. The figures
clearly show bouncing solutions for Linear, CPL, JBP, ASSS and
Efstathiou models with a minimal non-zero scale factor $a(t)$.\\

\begin{figure}
\includegraphics[height=1.6in]{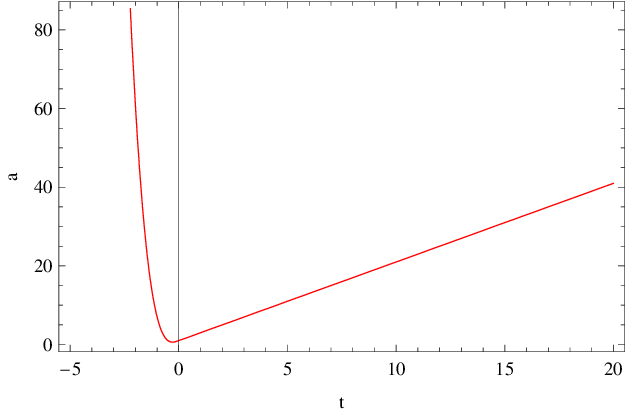}~~~~
\includegraphics[height=1.6in]{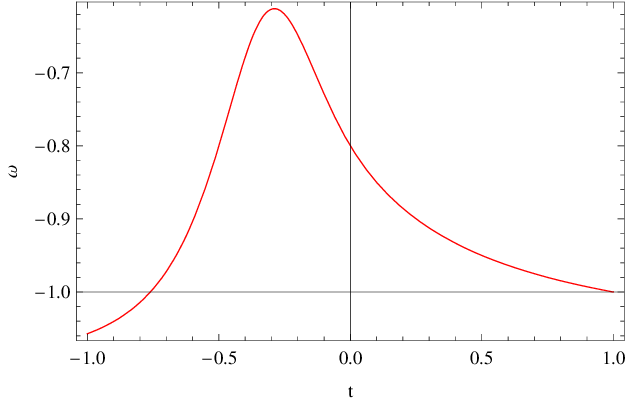}\\

\vspace{2mm} ~~~~~~~~~~~~Fig.36~~~~~~~~~~~~~~~~~~~~~~~~~~~~~~~~~~~~~~~~~~~Fig.37~~~~\\
\vspace{2mm}

\includegraphics[height=1.6in]{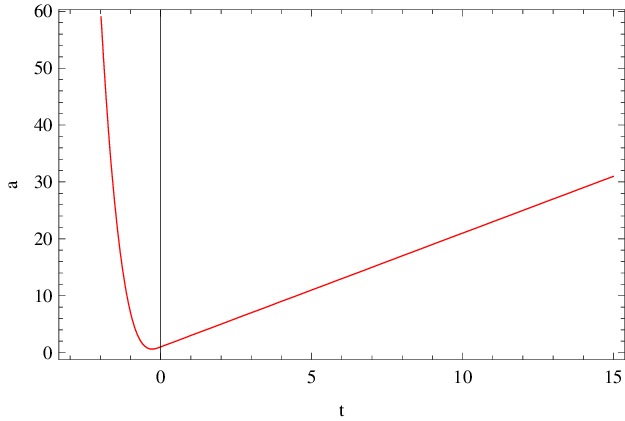}~~~~
\includegraphics[height=1.6in]{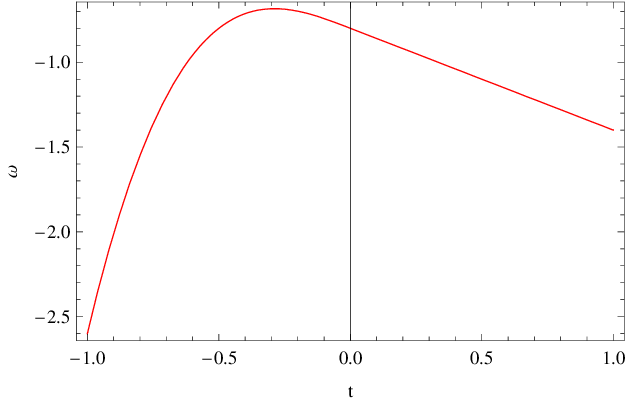}\\

\vspace{2mm} ~~~~~~~~~~~~Fig.38~~~~~~~~~~~~~~~~~~~~~~~~~~~~~~~~~~~~~~~~~~~Fig.39~~~~\\
\vspace{2mm}

\includegraphics[height=1.6in]{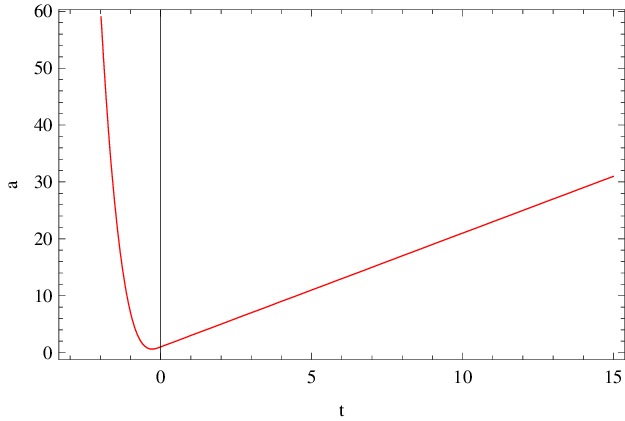}~~~~
\includegraphics[height=1.6in]{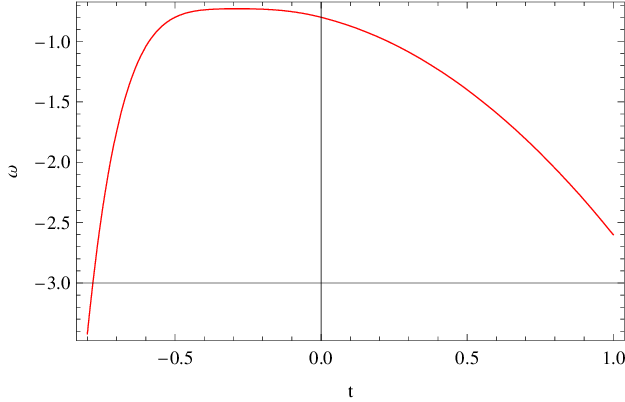}\\

\vspace{2mm} ~~~~~~~~~~~~Fig.40~~~~~~~~~~~~~~~~~~~~~~~~~~~~~~~~~~~~~~~~~~~Fig.41~~~~\\
\vspace{2mm}

\includegraphics[height=1.6in]{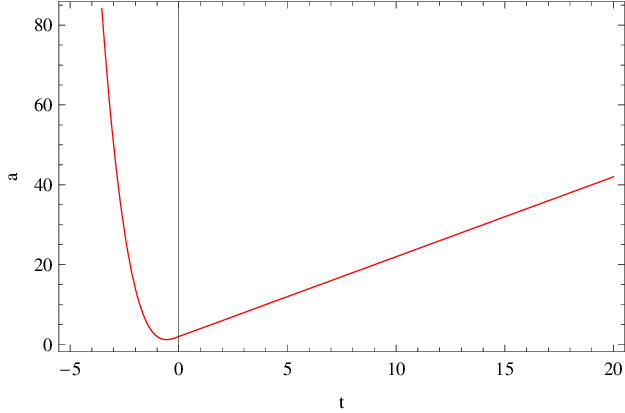}~~~~
\includegraphics[height=1.6in]{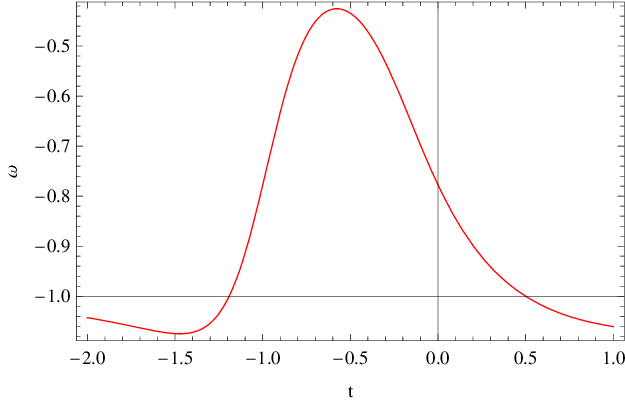}\\

\vspace{2mm} ~~~~~~~~~~~~Fig.42~~~~~~~~~~~~~~~~~~~~~~~~~~~~~~~~~~~~~~~~~~~Fig.43~~~~\\
\vspace{2mm}

\includegraphics[height=1.6in]{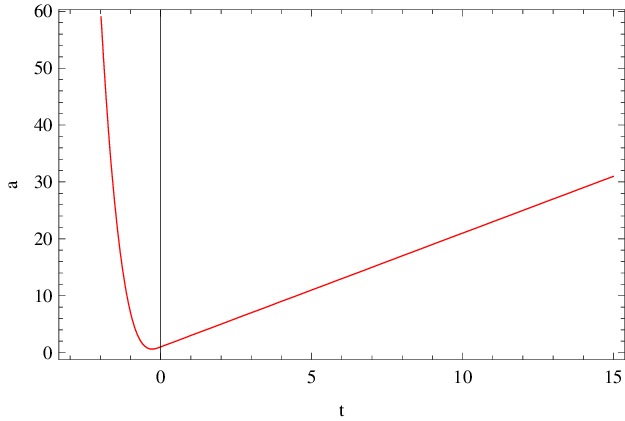}~~~~
\includegraphics[height=1.6in]{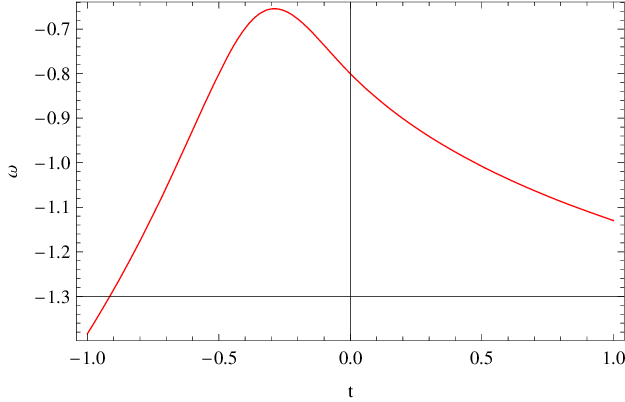}\\

\vspace{2mm} ~~~~~~~~~~~~Fig.44~~~~~~~~~~~~~~~~~~~~~~~~~~~~~~~~~~~~~~~~~~~Fig.45~~~~\\
\vspace{2mm}

Figs. 36-37 (Linear), 38-39 (CPL), 40-41 (JBP), 42-43 (ASSS) and
44-45 (Efstathiou) show the variation of scale factor $a(t)$
(left) and EoS parameter $\omega$ (right) with the variation in
cosmic time $t$ for the different parameterization models.\\

\vspace{6mm}

\end{figure}

\section{\normalsize\bf{Discussion}}

In this work, we have considered the D-dimensional flat FRW model
of the Universe in the background of some well known
parametrization of dark energy models like linear, CPL, JBP, ASSS
and Efstathiou. By considering the scalar field model as these
parametrizations of dark energy, we found the energy density,
pressure, scalar field and corresponding potential in terms of the
redshift $z$. In model I (linear), we have drawn the variations of
energy density $\rho$, pressure $p$, EoS parameter $\omega(z)$,
Hubble parameter $H$, potential $U(\phi)$ and the scalar field
$\phi$ with the variation in $z$ in Figs. 1 - 6 respectively. Fig.
7 shows the variation of $U(\phi)$ with $\phi$. We have seen that
$\rho$, $H$ and $\phi$ decrease as $z$ decreases. $p$ and $\omega$
decrease from positive level to negative level as $z$ decreases.
$U(\phi)$ first increases up to certain value of $z$ and then
decreases as $z$ decreases. Also $U(\phi)$ decreases as $\phi$
increases.\\

In model II (CPL), we have drawn the variations of energy density
$\rho$, pressure $p$, EoS parameter $\omega(z)$, Hubble parameter
$H$, potential $U(\phi)$ and the scalar field $\phi$ with the
variation in $z$ in Figs. 8 - 13 respectively. Fig. 14 shows the
variation of $U(\phi)$ with $\phi$. We have seen that $\rho$, $H$,
and $U(\phi)$ first decrease up to $z=0$ as $z$ decreases and then
sharply increase as $z$ decreases. But $p$ first increases up to
$z=0$ as $z$ decreases and then sharply decreases as $z$ decreases
but keeps the negative level. $\omega$ decrease as $z$ decreases
and always keeps negative sign. $\phi$ increases as $z$ decreases.
Also $U(\phi)$ decreases as $\phi$ increases.\\

In model III (JBP), we have drawn the variations of energy density
$\rho$, pressure $p$, EoS parameter $\omega(z)$, Hubble parameter
$H$, potential $U(\phi)$ and the scalar field $\phi$ with the
variation in $z$ in Figs. 15 - 20 respectively. Fig. 21 shows the
variation of $U(\phi)$ with $\phi$. We have seen that $\rho$, $H$,
and $U(\phi)$ first decrease up to $z=0$ as $z$ decreases and then
sharply increase as $z$ decreases. But $p$ first increases up to
$z=0$ as $z$ decreases and then sharply decreases as $z$ decreases
but keeps the negative level. $\omega$ first increases up to
certain value of $z$ and then sharply decreases but keeps negative
sign. $\phi$ decreases as $z$ decreases. Also $U(\phi)$ increases
as $\phi$ increases.\\

In model IV (ASSS), we have drawn the variations of energy density
$\rho$, pressure $p$, EoS parameter $\omega(z)$, Hubble parameter
$H$, potential $U(\phi)$ and the scalar field $\phi$ with the
variation in $z$ in Figs. 22 - 27 respectively. Fig. 28 shows the
variation of $U(\phi)$ with $\phi$. We have seen that $\rho$, $H$,
$\phi$ and $U(\phi)$ decrease as $z$ decreases. $p$ increases as
$z$ decreases but keeps negative sign. $\omega$ first decreases up
to certain value of $z$ and then increases as $z$ decreases. Also
$U(\phi)$ increases as $\phi$ increases.\\

In model V (Efstathiou), we have drawn the variations of energy
density $\rho$, pressure $p$, EoS parameter $\omega(z)$, Hubble
parameter $H$, potential $U(\phi)$ and the scalar field $\phi$
with the variation in $z$ in Figs. 29 - 34 respectively. Fig. 35
shows the variation of $U(\phi)$ with $\phi$. We have seen that
$\rho$, $H$, and $U(\phi)$ first decrease up to certain value of
$z$ as $z$ decreases and then sharply increase as $z$ decreases.
But $p$ first decreases from positive level to negative level then
slightly increases and then sharply decreases as $z$ decreases but
keeps the negative sign. $\omega$ decreases from positive level to
negative level as $z$ decreases. $\phi$ decreases as $z$
decreases. Also $U(\phi)$ increases as $\phi$ increases.\\

Also the present work is designed to investigate the possibilities
of finite time future singularities i.e., types I (big rip), II
(sudden), III and IV singularities. From the table, we observe
that for linear parametrization model, the type I (big rip)
singularity occurs provided $1+\omega_0-\omega_1<0$, but types
II-IV singularities cannot occur. For CPL parametrization model,
the type I (big rip) singularity occurs provided $\omega_1>0$, but
types II-IV singularities cannot occur. For JBP parametrization
model, the type I (big rip) singularity occurs provided
$\omega_1>0$, but types II-IV singularities cannot occur. For ASSS
parametrization model, the types I-III  singularities cannot
occur, only type IV singularity occurs provided ${A_1}^2\geq
4A_0A_2$. For Efstathiou parametrization model, the type I (big
rip) singularity occurs provided
$\omega_1>0$, but types II-IV singularities cannot occur.\\\\

{\bf Conflicts of Interest}: The authors declare that there is no
conflict of interest regarding the publication of this paper.\\\\

{\bf Acknowledgement}: The authors are thankful to IUCAA, Pune,
for their warm hospitality where most of the work has been done
during a visit under the Associateship Programme. \\\\

\end{document}